\documentclass[prb,balancelastpage,twocolumn,footinbib,superscriptaddress,showpacs]{revtex4-1}
\usepackage{color,amsthm,amsmath,amsfonts,graphicx,bm}
\usepackage{bm}
\usepackage{color}
\usepackage{amsfonts}
\usepackage{amssymb}
\usepackage{amsmath}
\usepackage{epsfig}
\usepackage{graphicx}
\usepackage{enumerate}
\usepackage{multirow}
\usepackage{verbatim}
\usepackage{color}
\usepackage{subfigure}

\begin{document}

\title{Loop update for infinite projected entangled-pair states in two spatial dimensions}

\author{Yi Zheng}
\author{Shuo Yang}
\email{shuoyang@tsinghua.edu.cn}
\affiliation{State Key Laboratory of Low-Dimensional Quantum Physics and Department of Physics, Tsinghua University, Beijing 100084, China}


\begin{abstract}
We propose an improved approach to carry out the imaginary time evolution of infinite projected entangled-pair states (iPEPS), especially for systems with criticality. A cyclic optimal truncation is introduced to update the tensors along a closed loop, aiming to remove the redundant internal correlations. We demonstrate the algorithm by considering an elaborate evolution based on simple update on a small plaquette. This scheme can also be applied to a full update strategy. We demonstrate their performances on simulating the ground states of the spin-$1/2$ anti-ferromagnetic Heisenberg model and the transverse field Ising model on a square lattice.
\end{abstract}

\maketitle

\section{Introduction}\label{TNS}
Tensor-network states (TNS) and the related methods have provided a versatile toolbox for studying the classical and quantum many-body systems\cite{Verstraete2008,Cirac2009,Zhao2010,Orus2014,Evenbly2015,Evenbly2017}. Some of the most prevailing TNS, such as matrix product states (MPS)\cite{Fannes1992,Vidal2006,Pippan2010,Schollwock2011}, projected entangled-pair states (PEPS)\cite{Verstraete2004,Jiang2008,Jordan2008,Orus2009,Corboz2010,Lubasch2014} and multi-scale entanglement renormalization ansatz (MERA)\cite{Vidal2008,Giovannetti2008,Corboz2009,Evenbly2015b}, are recently utilized to capture the entanglement features of states\cite{Verstraete2004b,Gu2008,Gu2009,Vidal2007}. For numerical applications, these TNS serve as the basis for variational approaches and real (imaginary) time evolutions\cite{Verstraete2004,Schollwock2011,Vidal2006,Vidal2007,Jiang2008,
Jordan2008,Orus2009,McCulloch2007,McCulloch2008,Crosswhite2008,Corboz2016}.

Most tensor network (TN) algorithms are in general dealing with the tradeoff between accuracy and efficiency. To find the approximate ground state of a local Hamiltonian, the simple update (SU), based on infinite time-evolving block decimation (iTEBD), counts the environment by effective weights from Schmidt decompositions. Such a strategy is optimal for the canonicalized TNS\cite{Vidal2006} and can be applied approximately to PEPS as well. Near criticality, however, the growth of truncation errors may defeat the efficiency, especially in two (or higher) dimensional gapless systems. The full update (FU), on the other hand, constitutes an accurate but time-consuming strategy by taking the full environment into account in each iteration\cite{Jordan2008,Orus2009,Corboz2016}. The insight is to capture the long-range correlations through a brute-force calculation. Several moderate algorithms have also been proposed, such as cluster update\cite{Wang2011} and fast FU\cite{Phien2015}. To implement and further improve these schemes, there are two main ingredients to be concerned: (i) the calculation of the effective environment, and (ii) an optimal local truncation. By formulating the environment as TN contractions, methods such as coarse-graining\cite{Levin2007,Zhao2010}, boundary MPS (BMPS)\cite{Jordan2008} and corner transfer matrix (CTM)\cite{Orus2009} provide feasible calculations with large computational costs. Further progresses include recycling the environment during TN updating\cite{Phien2015,Phien2015b} and the recent proposed nested TN formalism\cite{Xie2017}. For the second issue, the singular value decomposition (SVD) provides optimal truncations for the canonicalized MPS and tree-TNS. However, such a canonical form may not exist in PEPS or MERA since these TN contain closed loops. Recent works have also shed new light on possible canonicalization of PEPS\cite{Zaletel2019,Haghshenas2019}, though investigations of manipulating infinite PEPS (iPEPS) are still needed. As for the truncation of a local bond, there are some prominent candidates besides the variational MPS method\cite{Verstraete2008,Verstraete2004b}, such as full-environment optimal truncations\cite{Evenbly2018} and graph-independent local truncations\cite{Hauru2018}, based on the bond environment and the environment spectrum, respectively.

As the widely-adopted SU scheme may oversimplify the information about environments which surround the two updating sites, leading to states with inadequate accuracy near criticality. It is of fundamental interest to address the limitation and to search for algorithms that boost the accuracy without costing large efforts. Inspired by our previous study on loop tensor network renormalization (Loop-TNR)\cite{Yang2017}, here we present a simple proposal to improve the SU algorithm. A loop structure and a loop truncation scheme are introduced to rearrange and remove redundant correlations. We consider an elaborate simulation by evolving the state on a loop structure in each iteration. Such a loop update (LU) approach can yield more accurate results in comparison with the traditional SU and is computationally cheaper than FU. To make simulations in the thermodynamic limit, the spacial translational invariance is considered as in the iPEPS algorithms. The concept of LU provides a direct improvement for other TN algorithms including the FU scheme. Their performances are tested by simulating the ground states of the spin-$1/2$ anti-ferromagnetic Heisenberg model and the transverse-field Ising model on a 2D square lattice.

This paper is organized as follows. In Sec. \ref{LU}, we introduce the main idea of the loop update algorithm based on the implementation of SU. The performances are tested by simulating the gapless models in Sec. \ref{Models}. A summary is presented in Sec. \ref{summary}.

\section{Loop update algorithm}\label{LU}
We start by considering an iPEPS as the variational ansatz of a local Hamiltonian on a square lattice. Generalizations to other lattices are straightforward. The iPEPS is interpreted as a unit cell of $2\times2$ tensors $\Gamma_i[m_i]$ ($i = 1,2,3,4$), which is periodically repeated. Here $m_i$ denote the physical indices in Hilbert space of dimension $d$. We assume that the iPEPS wave function is given by
\begin{eqnarray}
_A\langle m_1,m_2,m_3,m_4|\Psi\rangle = \text{tTr}\prod_{\alpha,\beta,i}
\lambda_\alpha\lambda_\beta\Gamma_i^A[m_i]
\end{eqnarray}
where $\alpha\in\left\{u,r,d,l\right\}$ and  $\beta\in\left\{\bar{u},\bar{r},\bar{d},\bar{l}\right\}$ represent the links shown in Fig. 1(b). $\lambda_{\alpha(\beta)}$ are diagonal weight matrices attached to such links. The four tensors $\Gamma_i^A$ are located at the corners of a square plaquette labeled by $A$ [the darker gray square in Fig. 1(a,b)]. Here the multiplications indicate tensor contractions of virtual bonds within the plaquette, and the tensor trace is carried out for the square cluster. The Hamiltonian can be decomposed as $H = \sum h^{[A]}+h^{[B]}$, where $h^{[A(B)]}$ involves all local operators acting on the $A$ ($B$)-plaquette. We have the Suzuki-Trotter expansion $e^{-iHdt}\simeq e^{-ih^{A}dt}e^{-ih^{B}dt}+O(dt^2)$ in real time evolutions. The ground state searching scheme is achieved by replacing $idt$ with an imaginary time step $\delta\tau$. We recall that for iTEBD algorithm\cite{Vidal2006}, the evolution operator is applied as a local gate which bridges two neighboring sites. Instead, here $h^A$ ($h^B$) describes a 4-site system in the $A(B)$-loop. We express $U(h^{A(B)}) = e^{-ih^{A(B)}dt}$ as a set of matrix product operators (MPO) $U_i^{A(B)}$ ($i=1,2,3,4$) with virtual dimension $\chi_\text{mpo}$, as shown in Fig. 2(b). Thus the algorithm is achieved by alternatively applying those MPOs to $\Gamma_i^A$ and $\Gamma_i^B$. Note that the $A$ and $B$ blocks consist of the same tensors with different orders, and they act as an effective environment for each other. The cyclic structure in $A(B)$ is analogous to the loop construction in optimizing the tensor network renormalization (TNR)\cite{Yang2017}. The aim is to rearrange local entanglement according to the efficient MPS algorithms, leading to an improved accuracy for systems with criticality.

\begin{figure}
  \centering
  \includegraphics[width=8cm]{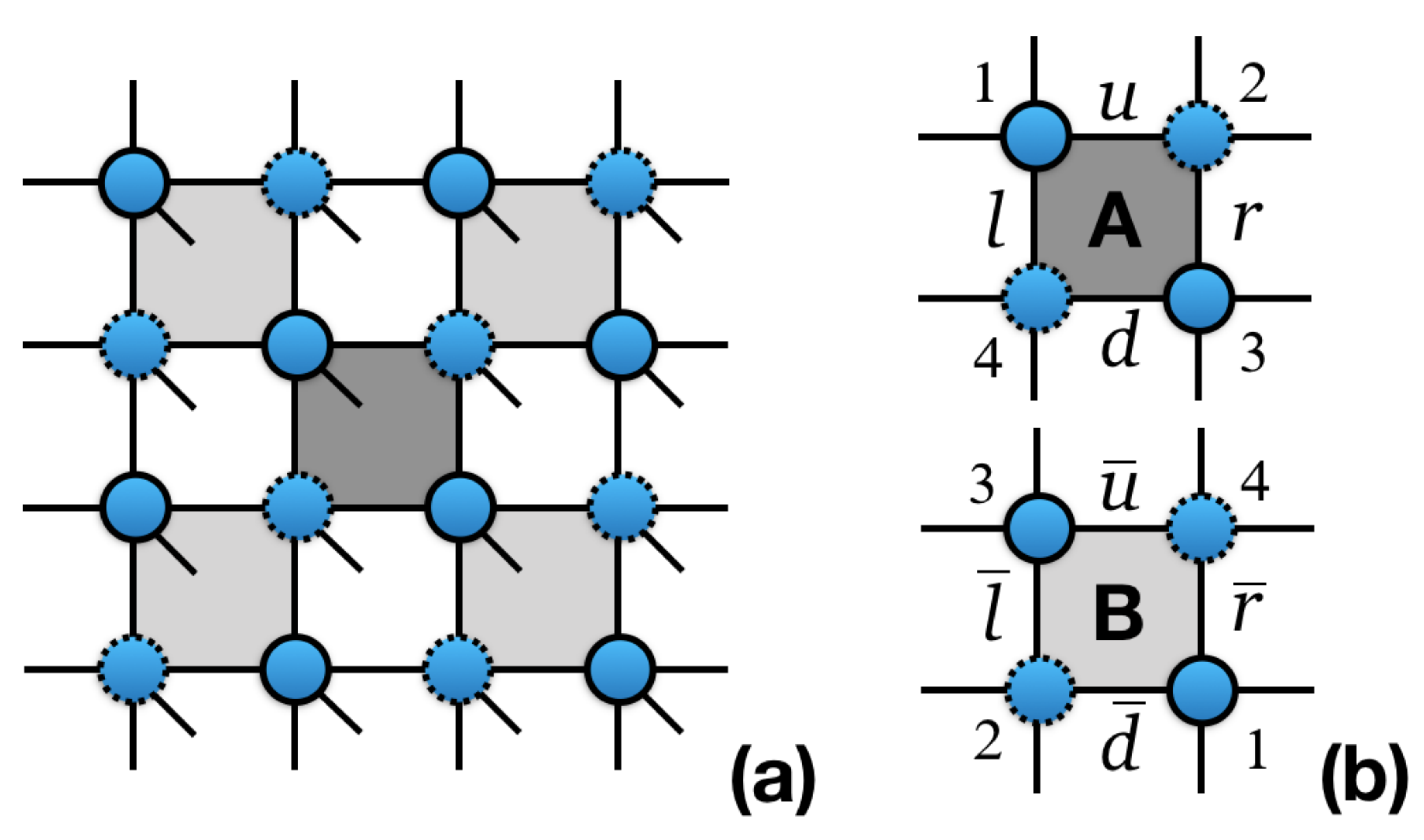}\\
  \caption{(a) Schematic representation of the TNS for the square lattice system. The lattice is divided into two parts formed by the darker and lighter shaded plaquettes. (b) The two plaquettes in (a), with the same local tensors but in different orders. The physical indices and the weight matrices $\lambda_{\alpha(\beta)}$ are not shown explicitly.} \label{Fig1}
\end{figure}

We demonstrate the related algorithm in the SU strategy. The first step is to evolve the state obtained from the previous iteration. Focusing on the local $A$-loop, the evolved state is expressed as $U(h^A)|\Psi\rangle$. We define the tensor cluster by
\begin{eqnarray}\label{TA}
\mathcal{F}_A[\tilde\lambda_\alpha,\tilde\Gamma_{i=1-4}^A] = \prod_{\alpha,i}
\tilde\lambda_\alpha\tilde\Gamma_i^A[m_i],
\end{eqnarray}
where $\tilde\lambda_{\alpha\in\left\{u,r,d,l\right\}}=\lambda_\alpha\otimes I$ are the enlarged weight matrices combined with $\lambda_\alpha$ and identity matrices. The local tensors are given by
\begin{eqnarray}
\tilde\Gamma_i^A[m_i] = \prod_\beta\lambda_\beta^2\sum_{m_i'} U_i^A[m_i,m_i']\Gamma_i^A[m_i'].
\end{eqnarray}
We have involved two weight matrices $\lambda_{\beta}$ on branches emitting outwards of the $A$-plaquette, as in the SU for both iTEBD\cite{Vidal2006} and iPEPS\cite{Jiang2008}. Thus, such matrices need to be returned at the end of each iteration by applying $\lambda_\beta^{-1}$.
Eq. \ref{TA} also implies an MPS with periodic boundary condition. The dimension of the virtual bonds is $D\chi_\text{mpo}$ with $D$ as a preset threshold for iPEPS. Each physical leg and two open branches are combined to give a bond dimension $D^2d$ as shown in Fig. 2(c).

The main task is to perform an accurate truncation through entanglement filtering and preserve the quasi-canonical form simultaneously. There are several options to achieve this goal: (i) As in the well-established variational methods for MPS\cite{Verstraete2008,Verstraete2004b}, the truncated tensors can be obtained by minimizing the cost function which corresponds to the distance between two MPSs. The computational cost scales as $O(D^6\chi_\text{mpo}^6)$. (ii) A quasi-canonical form is obtained through successive SVD processes and a final truncation\cite{Kalis2012}, though, the burden of computing is heavy [$\propto O(D^7\chi_\text{mpo}^7d^2)$]. (iii) The typical canonicalization procedure\cite{Orus2008}, which is designed for MPS under non-unitary evolutions, prefers the translation invariant condition instead of the cyclic case. However, this method requires a lower computational cost which scales as $O(D^5\chi_\text{mpo}^3d)$. (iv) The full environment truncation (FET)\cite{Evenbly2018}, an optimal candidate for TN with closed loops, is based on the choice of local gauge freedom. In this method, two isometries $\mu$ and $\nu$ are determined according to $\tilde\lambda_\alpha\sim\mu\lambda_\alpha'\nu^\dag$ with $\lambda_\alpha'$ of dimension $D$. The so-called bond environment for bond-$\alpha$ ($\alpha\in\left\{u,r,d,l\right\}$) is counted in variational techniques. The leading computational cost is the same as that in (i).

\begin{figure}
  \centering
  \includegraphics[width=8cm]{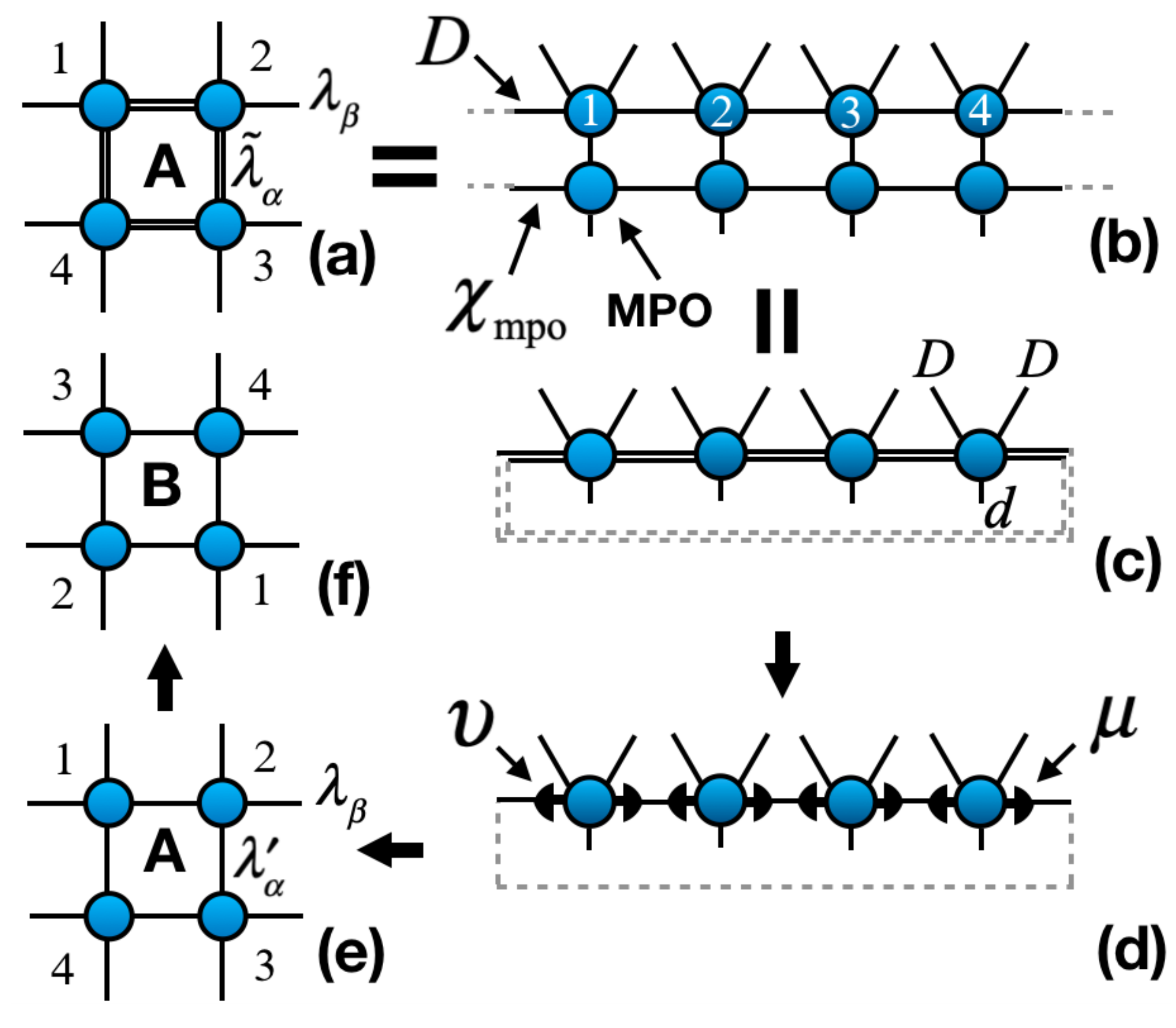}\\
  \caption{Schematic diagram of cyclic optimal truncations in the updating algorithm. (a-c) Equivalent representation of the loop configuration with enlarged bond dimensions ($D\chi_\text{mpo}$). The MPO formula of the evolution operator is applied to the $A$-loop. The skew legs in (b,c) are open legs in (a). The weight matrices $\tilde\lambda_\alpha$ and $\lambda_\beta$ reside in virtual bonds and open legs, respectively. Typical bond dimensions are labeled. (c-e) The cyclic operation of the FET, where the weight matrices $\lambda_\alpha'$ and the projectors $\mu$ and $\nu$ are variationally updated. (f) The $B$-loop is obtained by rearranging the order of the renewed local tensors in (e).} \label{Fig1}
\end{figure}

Here, a pre-optimization is carried out by performing the canonicalization in (iii) and the form of Eq. \ref{TA} is retained. This can be significantly accurate since the dimension of each virtual bond is preserved. Sequentially, we adopt the last option as the truncation scheme. The Schmidt weight matrix and two isometries for a particular bond are initialized by $\tilde\lambda_\alpha \overset{\text{SVD}}{=}\mu\lambda_\alpha'\nu^\dag$. They are variationally updated by maximizing the fidelity $F = \langle f|i\rangle\langle i|f\rangle/(\langle f|f\rangle\langle i|i\rangle)$, here $|i\rangle$ and $|f\rangle$ are the initial and final state diagrammatically illustrated in Fig. 2(a) and (e), respectively. $\mu$ and $\nu$ then serve as projectors which reduce the bond dimension from $D\chi_\text{mpo}$ to $D$. We arrive at $\mathcal{F}_A[\lambda_\alpha',\Gamma_i']$ with $\Gamma_i' = \nu^\dag\tilde\Gamma_i\mu$, as illustrated in Fig. 2(e). To accomplish one iteration, the above procedures are carried out for the $B$-plaquette which is constructed by tensor switching [Fig. 2(e)$\rightarrow$(f)].

In addition, our algorithm is also applicable to the FU strategy. The effective environment for plaquette $A$($B$) is obtained through the well-known approximate methods such as BMPS\cite{Jordan2008} and CTM\cite{Orus2009,Nishino1996,Corboz2016}. In each evolution step, the main cost of calculating the environment is $O(D^6\chi^3)$ with $\chi>D^2$ the two-layer truncated bond dimension. The resulting environment is also necessary for the calculations of physical expectations and correlations.

\section{Heisenberg model and transverse field Ising model}\label{Models}
We benchmark the proposed LU algorithm in typical models. First, we consider a gapless spin-$1/2$ anti-ferromagnetic Heisenberg model on a square lattice: $H = \sum_{\langle i,j\rangle}h_{ij}$, with $h_{ij} = \mathbf{S}_i\cdot\mathbf{S}_j$ and $\mathbf{S}_i=(\sigma_{i}^{x},\sigma_{i}^{y},\sigma_{i}^{z})/2$. Such a model generates the starting point to understand various complex magnetic structures in solids. The quantum Monte Carlo (QMC) gives the ground state energy per bond $E_0 = -0.334719$ [$-0.669437(5)$ per site] and the staggered magnetization $M_0 = 0.30703$ for a $16\times16$ system\cite{Sandvik1997}.

\begin{figure}
  \centering
  \includegraphics[width=8.5cm]{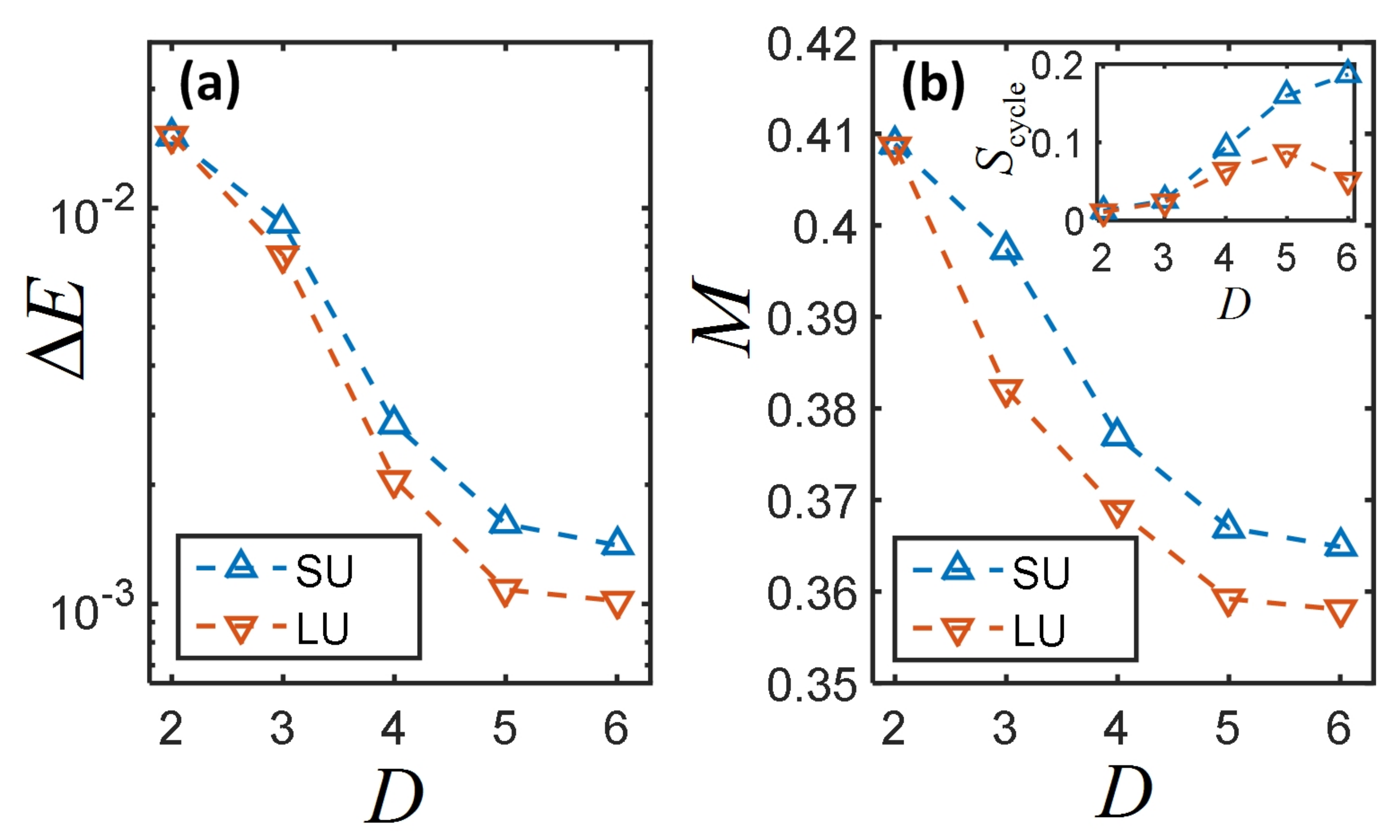}\\
  \caption{Simulation results of the Heisenberg model. (a) The relative error $\Delta E=(E_0-E)/E_0$ versus the dimension $D$ of the virtual bond. (b) The corresponding values of the staggered magnetization. The iPEPSs are obtained using loop update and simple update. Dashed lines are a guide to the eye. Inset: the cycle entropy $S_\text{cycle}$ of the iPEPSs from both algorithms.} \label{Fig1}
\end{figure}

In this model, the local MPO of the evolution operator can be constructed from the local gate $e^{-\delta\tau h_{ij}}$ that connects two neighboring spins\cite{Kalis2012}. The sequence of applying the gate has a negligible influence on the final results. Another option is to simply approximate $U_i$ by\cite{Zaletel2015}
\begin{eqnarray}
U(\delta\tau)=\begin{bmatrix}
   \mathbf{I} & -\delta\tau S^{x} & -\delta\tau S^{y} & -\delta\tau S^{z} \\
   S^{x}        & 0       & 0       & 0       \\
   S^{y}        & 0       & 0       & 0       \\
   S^{z}        & 0       & 0       & 0
   \end{bmatrix},
\end{eqnarray}
which is accurate to $O(\delta\tau)$ under periodic boundary conditions. One needs to start the iteration with a small time step $\delta\tau$ and gradually reduce $\delta\tau$ until convergence. The two constructions of MPO, both with $\chi_\text{MPO}=4$, are equivalent as long as the final $\delta\tau$ is small enough (\textit{e.g.}, to $\sim10^{-5}$).

In Fig. 3, we demonstrate the performance of the LU, compared with SU where the truncations are performed with SVD. As expected, in Fig. 3(a), the accuracy is controlled by the dimension $D$ of the virtual bonds. We get $E_\text{LU}\simeq-0.334377$ and $E_\text{SU} \simeq -0.334247$ with $D=6$, which agree with the QMC estimations within $0.102\%$ and $0.141\%$, respectively. Note that the relative error $\Delta E$ can be reduced efficiently through imaginary time evolutions. A smaller $\Delta E$ appears for the LU algorithm that contains cyclic optimization. As shown in Fig. 3(b), the staggered magnetization is also reduced in LU. We emphasize that the insight of LU is to remove redundant information through optimal truncations along closed loops. Specifically, we introduce the cycle entropy $S_\text{cycle}$\cite{Evenbly2018} as a measure of the internal correlations in those loops. The cycle entropies are calculated with the TNS from both algorithms. For $D = 2$ and $3$, we get the decrement from SU to LU: $S_\text{cycle} = 0.0124\to0.0105$ and $0.0256\to0.0235$. The amount of reduction grows with bond dimension $D$ as shown in the inset of Fig. 3(b). We achieve a reduced $S_\text{cycle}$ in LU since the FET is designed to remove irrelevant internal correlations. This is in the same spirit of removing short-range entanglement in Loop-TNR\cite{Yang2017,Gu2009}.

\begin{figure}
  \centering
  \includegraphics[width=8.6cm]{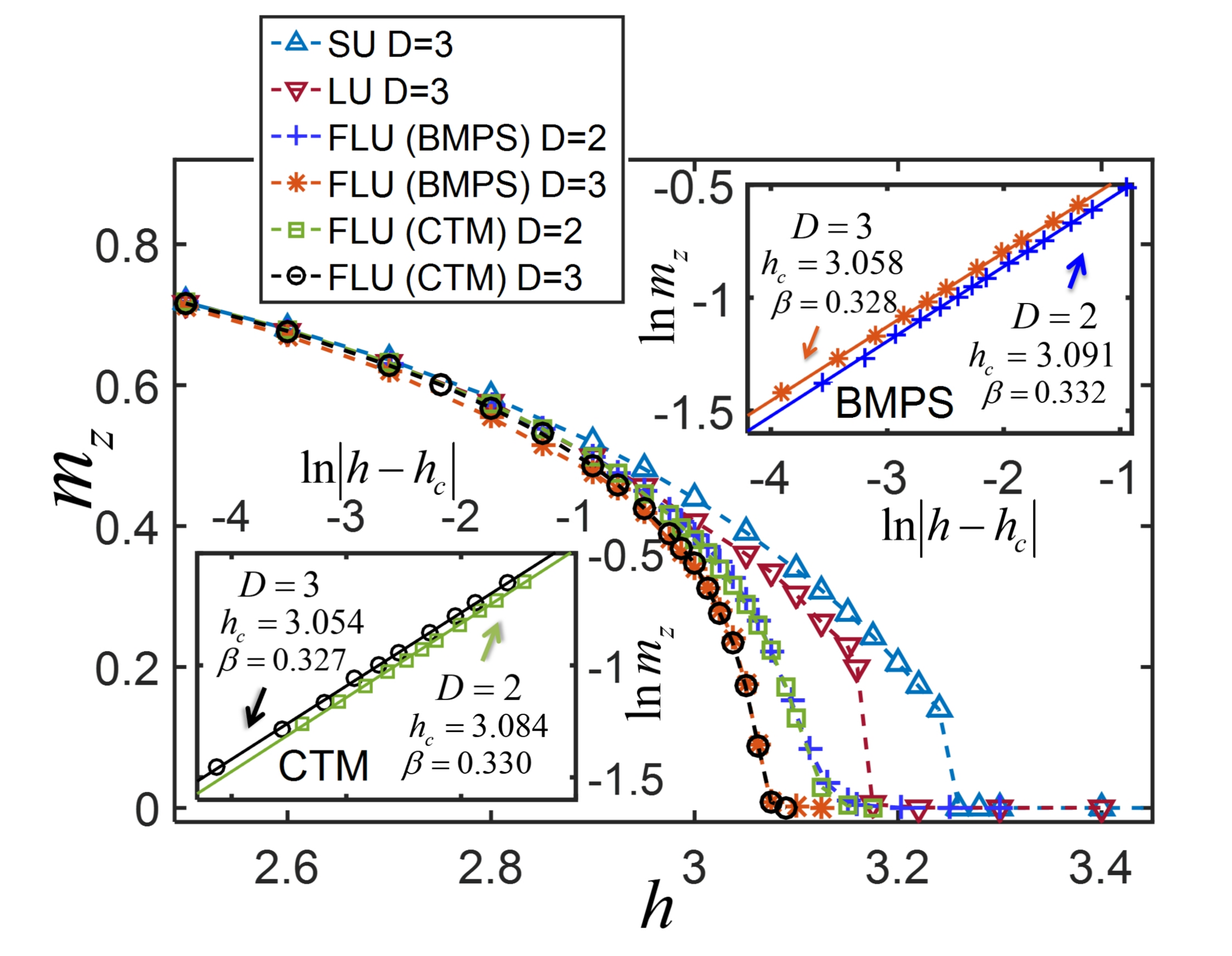}\\
  \caption{The magnetization $m_z$ as a function of the transverse field $h$, with dashed lines a guide to the eye. Insets: log plots of $m_z$ versus $|h-h_c|$ for the FLU simulations with the environments calculated from the BMPS and the CTM methods for $D=2$ and $3$. The results are compared with linear fits. The estimated critical fields and critical exponents are labeled in insets. } \label{Fig1}
\end{figure}

As another example, we consider the transverse-field Ising model $H = -\sum_{\langle i,j\rangle}\sigma_{i}^{z}\sigma_{j}^{z}-h\sum_i\sigma_{i}^{x}$ on a square lattice. The MPO with $\chi_\text{MPO}=2$ can be expressed by
\begin{eqnarray}\label{Ising}
U(h,\delta\tau)=\begin{bmatrix}
   \mathbf{I}+ h\delta\tau\sigma^{x} & \delta\tau\sigma^{z} \\
   \sigma^{z}        & 0
   \end{bmatrix},
\end{eqnarray}
which is again accurate to $O(\delta\tau)$. We simulate the ground states with different values of $h$. The magnetization $m_z=|\langle\sigma^{z}\rangle|$ is calculated as the order parameter. From QMC estimates, the phase transition occurs at the critical field $h_c^\text{QMC}\simeq3.044$, and the critical exponent for $m_z$ is given by $\beta^\text{QMC}\simeq0.327$\cite{Blote2002}. The results of $m_z$ versus $h$ are exhibited in Fig. 4. Here, the LU and SU with $D=3$ are compared with the full loop-update (FLU) which takes the effective environment into account when applying LU. Under off-critical conditions, we get close results in all these simulations. For $h=2.6$, the differences of $m_z$ are less than $10^{-2}$ and the energy offsets are less than $10^{-3}$. Approaching $h_c$, the deviations increase since the major long-range correlation starts to play an important role. Compared with SU, partial improvement appears in LU, which indicates the effectiveness of the cyclic optimal truncations.

In the FU simulations for models that are away from quantum critical points, the BMPS and CTM approaches are nearly equivalent in determining the two-site environments. Here we implement the FLU with these methods which generate the effective environments for the plaquettes of Fig. 1(b). The magnetizations with FLU for $D=2$ and $3$ are shown in Fig. 4, demonstrating a better-characterized quantum phase transition. The values of the critical fields $h_c$ and the critical exponents are illustrated in log plots, see insets of Fig. 4. Our estimations compared with the traditional FU results are given by:\\

\begin{tabular}{p{3cm}p{1.2cm}p{1.2cm}|p{1.2cm}p{1.2cm}}
               & $h_c^{D=2}$ & $\beta^{D=2}$ & $h_c^{D=3}$ & $\beta^{D=3}$ \\
FU with BMPS\cite{Jordan2008}:  & 3.10      & 0.346        & 3.06       & 0.332 \\
FU with CTM\cite{Orus2009}:  & 3.08      & 0.333        & 3.04       & 0.328 \\
FLU with BMPS: & 3.091     & 0.332        & 3.058      & 0.328\\
FLU with CTM: & 3.084     & 0.330        & 3.054      & 0.327\\
\end{tabular}\\

According to the QMC results, the above comparisons indicate a significant improvement of FLU in quantifying critical exponents, especially for $D=2$. For near-critical systems, methods with CTM are believed to be more suitable for simulations with inadequate bond dimensions $D$, as in the case of FU\cite{Orus2009}. We arrive at very similar results in the FLU with both environment calculation methods. Improved results are obtained in approaches with LU, which is a consequence of the cyclic truncations in loop-updating procedures.

Note that LU prevails over the $2\times2$ cluster update\cite{Wang2011}. This concept can be further extended to cluster update with a larger cluster size. As iPEPS is well suited to express the ground state ansatz of a local Hamiltonian, these iPEPS approaches show prominent abilities to study quantum phase transitions and critical phenomena even with a small bond dimension.

There are further constructive operations that are applicable to these algorithms. In each iteration of LU, the effective environments (\textit{i.e.}, the local weight matrices) are obtained from the cyclic truncations in the previous update. The BMPS and CTM environments can be recycled in a similar manner, and thus the fast full update and gauge fixing scheme in Ref. \onlinecite{Phien2015} may be favorable to the efficiency. In addition, both BMPS and CTM are dealing with tensor contractions of a two-layer TN. This can be converted to single-layer tensor contractions by implementing the nested TN method\cite{Xie2017}, which is instructive in reducing the computational cost and memory consumption.

\section{Summary}\label{summary}
We have proposed the LU algorithm of an iPEPS based on the cyclic FET that makes an optimal truncation. The benefits of such loop optimization are to remove redundant internal correlations and to provide more accurate results for critical systems. The LU is further upgraded by considering the full environment, which constitutes the FLU scheme. We demonstrate their performances with the spin-$1/2$ anti-ferromagnetic Heisenberg model and the transverse-field Ising model on a square lattice. The comparisons of these results indicate that the presented LU and FLU are the improved versions of SU and FU. Integrating 1D algorithms with 2D algorithms constitutes an important step to improve the performance of simulations. For future works, the LU scheme can be generalized to improve other TN simulations, including the real-time evolution, TNR in 3D, and finite-temperature evaluations.

\section{Acknowledgement}
S. Yang thank Zheng-Cheng Gu and Frank Pollmann for earlier collaborations. This work is supported by NSFC (Grant No. 11804181), the National Key R\&D Program of China (Grant No. 2018YFA0306504), and the Research Fund Program of the State Key Laboratory of Low-Dimensional Quantum Physics (Grant No. ZZ201803).

\bibliography{Square_loop}

\end{document}